# Optimal National Policies towards Multinationals when Local Regions Can Choose between Firm-Specific and Non-Firm-Specific Policies

**Osiris J. Parcero***

This paper looks at a country's optimal central-government optimal policy in a setting where its two identical local jurisdictions compete to attract footloose multinationals to their sites, and where the considered multinationals strictly prefer this country to the rest of the world. For the sake of realism the model allows the local jurisdictions to choose between firm-specific and non-firm-specific policies. We show that the implementation of the jurisdictional firm-specific policy is weakly welfare dominant. Hence the frequent calls for the central government to ban the former type of policies go against the advice of this paper.

## 1. Introduction

This paper looks at a country's optimal central-government policy in a setting where two of its local jurisdictions compete to attract footloose multinationals (hereafter MNCs) to their sites. Despite its increasing relevance, the joint consideration of the roles played by both the central government and local jurisdictions in this competition process has been neglected:

"[I]t is also important to incorporate sub-national governments into the competitive framework. In federal nations and large countries with decentralised administration it is often sub-national governments that deliver incentive packages and contribute most to both intra-national and international competition . . . (Charlton, 2003, p. 15)."[1]

* Kazakh-British Technical University, 59 Tole Bi Street, Almaty 050000 Kazakhstan (osirisjorge.parcero@gmail.com). I would like to thank Paul Grout, James R. Hines Jr., Oliver Kirchkamp, Carlos Ponce, Helmut Rainer, Tobias Regner, Wendelin Schnedler, Mariano Selvaggi, Michael Whinston, Abdul Aziz, O. Abahindy, and Abdul Rashid Dad for valuable discussions.

1 For similar expressions and related evidence see Mittermaier (2009) and Crabbé (2013).

One widespread feature of this competition process is that subnational governments often implement firm-specific policies; this is despite many voices against them (see Van Biesebroeck, 2010). For instance, among U.S. states and cities "bidding wars" for footloose plants have intensified since the 1960s, with incentive packages escalating in total worth (LeRoy, 2005; Chirinko and Wilson, 2008). Moreover, since the early 1990s, the same type of subnational competition has begun to proliferate in developing countries such as Brazil (Versano et al., 2002), China (Xu and Yeh, 2005), and India (Schneider, 2004), to mention only a few.

Another stylized fact is that countries' central governments often favor some locations in order to give them advantages at the time of competing for footloose MNCs – as modeled in Davies and Ellis (2007) and Parcero (2007). This asymmetric treatment may take the form of special economic zones (SEZs) created in particular regions or jurisdictions, which have been widely used by many countries in order to both attract FDI and promote exports. The privileges that these SEZs enjoy give them significant competitive advantages relative to other, nonfavored locations of a country. In the words of Easson:

"Many countries around the world provide tax incentives to establish investments in particular locations or regions within the country – from the coastal cities and special economic zones in China to the enterprise zones and renaissance zones of the United Kingdom and the United States." (2001, p. 10)

MNCs often try to find the best location for their production plants. However, when suitable options are not abundant, they end up having a preference for a specific country, where they can obtain locational rents. Locational rents may result from the MNC's desire to get involved in profit-shifting activities, the country's good strategic location, or geographically localized intrafirm, interfirm, or technological spillovers (Caves, 1996; Görg and Strobl, 2003; Thomas, 2011; Loretz and Mokkas, 2015). Locational rents can be defined as the difference in the profits MNCs can obtain by locating in one country vis-à-vis locating elsewhere. Table 1 makes a clear distinction between profits and locational rents for five different MNCs. It shows that they have different outside options when considering the establishment of a production plant in a country.

In this paper the tax base is the MNCs' locational rents. Notice that a profit tax rate would be an appropriate tool to extract locational rents if they are identical to profits (as in table 1's cases B and F). However, it would not be an appropriate one when they are not identical. For instance, table 1 shows that a profit tax rate slightly higher than 0 % would not attract MNC A, but it would be an extremely low tax rate in terms of extracting the locational rents of MNC B. We can cite two separate pieces of related literature. First, there

**Table 1**
*Difference between Locational Rents and Profits*

| MNC | Profits if locating in the host country (before tax) (1) | Profits if locating abroad (after tax) (2) | Locational rents (1)–(2) | Optimal profit tax rate to attract the MNC* |
|---|---|---|---|---|
| A | 40 | 40 | 0 | 0 % |
| B | 40 | 0 | 40 | 100 % |
| C | 70 | 30 | 40 | 57 % |
| D | 140 | 100 | 40 | 29 % |
| E | 1040 | 1000 | 40 | 4 % |
| F | 70 | 0 | 70 | 100 % |

Note: *This is the profit tax rate that would leave the MNC indifferent between coming to the country and going abroad.

is a literature modeling interregional (or international) competition for footloose multinationals; see Dembour (2008) for a survey. For example, Bond and Samuelson (1986) model the fact that the tax competition between countries can take the form of a tax holiday. Behrens and Picard (2008) present a model in which governments bid for firms by taxing/subsidizing setup costs and in which the firms choose both the number and the location of the plants they operate. Menezes (2003) uses an auction framework to examine jurisdictional competition where the value of the new plant may differ among jurisdictions and depends on the private information held by each jurisdiction. Haufler and Wooton (2006) consider unilateral and coordinated tax policy in a union of two regions (A and B) that competes with a foreign potential-host region (C) for the location of a monopolistic firm. Matsumoto and Ohkawa (2010) extend this analysis to the case where there are both bidding and non-bidding regions. However, unlike our paper, in this literature there is no central government intervening in the competition between the lower-level jurisdictions.

Second, our paper is related to the public-finance literature on *concurrent taxation*, which looks at the case where different levels of governments independently set taxes on a common tax base. The framework used is the standard tax competition model of Zodrow and Mieszkowski (1986); see Madiès et al. (2004) for further discussion and references.

One difference between the public-finance literature on tax competition and the aforementioned literature on competition for foreign direct investment is that the latter deals with markets of imperfect competition while the former assumes perfect competition. Second, in the public-finance literature the competition is for capital rather than for firms facing discrete location choices. Only the second type of competition allows the implementation of firm-specific policies, which is a central issue in our research. A third difference is that in the public-finance literature the interjurisdictional competition for firms occurs in a closed-economy setting; exceptions are Janeba and Wilson (1999) and Davies (2005). Janeba and Wilson allow for commodity trade and capital mobility between the competing regions and the rest of the world, assuming that the country's central government sets the level of tariff protection.

Two papers combine features of the two mentioned literatures. First, Davies and Ellis (2007) build a model where local jurisdictions compete for multinationals in both taxes (through a second price auction) and performance requirements (PRs). Their model shows that the setting of central-government PRs that lower the surplus generated in the second-highest-bid jurisdiction are beneficial for the country; they ultimately result in the winning jurisdiction retaining a greater proportion of any generated surplus. Second, Parcero (2007) looks at the concurrent-taxation problem in a setting where two identical local governments bargain with a footloose multinational about the locational tax to be charged, while the first-moving central government of the country has to set the locational tax to be paid by the multinational in each of the two local jurisdictions. Similarly to Davies and Ellis (2007), Parcero also finds that it is optimal for the central government to give privileges to one jurisdiction (the favored one) relative to another (nonfavored) jurisdiction in order to increase its bargaining power vis-à-vis the multinational.

Finally, a number of papers have considered foreign-firm-specific taxes. One interesting recent application in the context of subnational government competition for FDI is Amerighi and De Feo (2014). In a setting where there is no difference between profits and locational rents (as in table 1's alternatives B and F), they analyze tax competition for a MNC between two asymmetric countries (local jurisdictions) of a region. One country provides a higher pre-tax locational rent to the MNC, while the other has the advantage of having a lower statutory profit tax rate. They consider two scenarios, one in which firms can exploit differences in statutory taxes by shifting profits to lower-tax jurisdictions, and another where they cannot.

Our paper differs from Amerighi and De Feo (2014) in four main ways. First, it allows profits to differ from locational rents, and so it assesses the optimality of locational rather than profit taxes. Second, it considers identical jurisdictions in contrast to the two-dimensionally differentiated ones in their paper. Third, it allows the local jurisdictions to choose between firm-specific and non-firm-specific policies; by doing so it also captures a distinctive feature of the former – viz., the existence of a negotiation with the MNC. Fourth, it incorporates a central government as an important player in the tax game. This allows comparison of a country's welfare under three different scenarios: cen-tral government only, local jurisdictions only, and both central governments and local jurisdictions.

We adopt Parcero (2007)'s framework where the central government of the host country moves first and sets (*tax-posts*) the lump-sum central government locational taxes (fees) to be paid by the footloose multinationals in the different locations. However, for the sake of realism we allow more flexibility to the local jurisdictions at the time of choosing their policies towards multinationals – namely, whether they are firm-specific or non-firm-specific policies.

It is known that firm-specific policies come in different degrees and modes; see Fisher and Peters (1999) for a more detailed categorization. However, in the present paper we will concentrate on how good these policies are in terms of taxing the locational rents obtained by footloose multinationals. Thus, the present paper is about tax competition and excludes any competition in terms of, say, infrastructure provision or regulation.

As proxies for the firm-specific and the non-firm-specific policies we use what we call the *tax-bargaining* and *tax-posting* regimes, respectively.[2] In the former regime each multinational negotiates the particular lump-sum jurisdictional locational tax to be paid; hence tax discrimination is the advantage of this regime. On the contrary, in the tax-posting regime the nonnegotiable lump-sum jurisdictional locational taxes have to be set in advance; this commitment to nonnegotiation provides a higher bargaining power to the tax-poster, though tax discrimination is not possible under this regime.[3]

The following results are found. First, as in Parcero (2007), we find that a central government's asymmetric tax treatment of the two identical jurisdictions welfare-dominates its symmetric one. Second, under some parameter constellations, prevention of competition (a prohibitively high central-government locational tax in the nonfavored jurisdiction) is not desirable. This result is a consequence of the existence of an out-of-equilibrium conflict of interest between the favored jurisdiction and the central government. In particular, when this prohibitively high locational tax is set in the nonfavored jurisdiction, the favored jurisdiction finds it optimal to implement the tax-posting regime, whereas the country would prefer it to choose the tax-bargaining one. This out-of-equilibrium conflict of interest is resolved by the existence of some competition from the nonfavored jurisdiction, which drives the favored jurisdiction to adopt the tax-bargaining regime.

Third, this paper recommends that local jurisdictions should be allowed to negotiate taxes with the multinationals. The reason for this result is as follows.

2 A similar classification is used in a subfield of industrial organization dealing with *price-bargaining* and *price-posting*.

3 This is a simplifying assumption based on the fact that bargaining allows higher price discrimination than price-posting; see Spier (1990) and Bester (1993).

When the favored jurisdiction tax-posts, the country as a whole is implementing a *tax-posting regime* (i.e., the sum of the two levels of government's posted taxes is equivalent to one posted tax). On the contrary, when the favored jurisdiction implements bargaining, the country as a whole is implementing a *bargaining-with-reservation-tax* regime (the reservation tax being the central government's locational tax in the favored jurisdiction), which, under the optimal central-government locational tax, provides a higher country welfare (aggregate revenue) than a tax-posting regime.[4]

If it turns out that our model is an appropriate simplification of reality, this last result might be utilized to refute those criticisms of interjurisdictional tax competition that are specifically addressed to the jurisdictional implementation of firm-specific policies. Hence, the frequent calls for the central government to ban this type of jurisdictional policies would go against the advice of this paper.

The structure of the paper is as follows. The basic model, which consists of a four-stage game, is introduced in section 2. In section 3 we look at the equilibrium of the three subgames, where the jurisdictional locational taxes are determined (stages 3 and 4 of the game). At these stages both jurisdictions have already chosen their tax regimes (in stage 2) and they know the central-government locational tax in each jurisdiction (set in stage 1). In section 3.1 we solve the subgame where one jurisdiction chooses the tax-bargaining regime and the other chooses the tax-posting one (there are two symmetric cases here). In section 3.2 we solve the subgame where both jurisdictions choose the tax-posting regime. In section 3.3 we solve the subgame where both jurisdictions choose the tax-bargaining regime. Section 4.1 solves the first stage of the game, where the optimal central-government locational tax in each jurisdiction has to be found. Section 4.2 does a country's welfare comparison under different scenarios of tax-related jurisdictional autonomy as well as complete centralization. Section 4.3 discusses the similarities and differences with the literature. Section 5 concludes and suggests possible lines for future research.

## 2. The Basic Model

We assume a four-stage game involving the central government $G$, two identical local jurisdictions, and a multinational $M_i$, $i = l,h$.[5] Here $M_h$ and $M_l$ show up with probabilities $q$ and $1-q$ respectively. In the absence of any cen-

4 In a different context, this bargaining regime is referred to by Wang (1995) as "bargaining with reservation price." He shows that it revenue-dominates a "simple bargaining" as well as a "posted" price.

5 The results of the paper would not be affected by considering more than one multinational.

tral government and jurisdictional taxes, $M_i$ would strictly prefer this country to the rest of the world; hence if $M_i$ locates in one of the jurisdictions, it obtains a locational rent $r_i$, with $r_h > r_l > 0$. In other words, the high-type MNC is one that produces high locational rents, and the low-type MNC is one that produces low locational rents. Assuming different types of MNCs captures the fact that they produce different locational rents, which in our paper are allowed to be different to profits. This is a very realistic feature that only by chance would not apply.[6] It is clear that, while $M_i$'s profit in a country may be verifiable (in a court of law), this is much less the case for the locational rents (i.e., the difference in the profits that $M_i$ obtains in different locations); hence, we assume they are nonverifiable.[7] Nonverifiability is relevant if MNCs produce different locational externalities, the tax base is the locational rents, and, as in the tax-posting regime, taxes have to be set in advance – in this scenario, tax discrimination becomes infeasible. Thus, these three features are needed in order to have a meaningful distinction between the tax-posting and the bargaining regimes. Finally, for simplicity and also because of the nonexistence of suitable asymmetric-information bargaining games we assume that all players have complete information at the time of making their decisions.

The sequence of the game is shown in figure 1. In the first stage of the game and with the aim of maximizing the country's expected welfare, $G$ posts a set of lump-sum locational taxes[8], $g_1$ and $g_2$ ($g_1 \leq g_2$), to be paid by $M_i$ in the case of locating in jurisdiction $J_1$ or $J_2$ respectively.[9] From now on we refer to $J_1$ and $J_2$ as the "favored jurisdiction" and the "nonfavored jurisdiction," respectively, even though, strictly speaking, this will be the case only if $g_1 < g_2$.

6 For expositional convenience we define $M_i$'s locational rent relatively to what it could get abroad, ignoring the comparison between the two jurisdictions. Moreover, for simplicity we assume that $M_i$ does not produce externalities.

7 Verifiability would not be a problem if the law stipulated (in advance) that a MNC has to pay a tax, say, equal to the entire profits it produces if attracted to a local jurisdiction. *Ex post*, the realized profits can be verified by the court, if needed. However, if the law stipulates (in advance) that the tax to be paid is equal to the MNC's locational rent, a court of law would not be able to verify it. In the latter case, verifiability would require that the court know the level of profits the MNC would have obtained if located in another country (something that clearly did not happen). The examples in table 1 should help to understand this point.

8 For simplicity, hereafter we refer to "tax" instead of "locational tax."

9 Notice that when the taxes are posted, the tax-poster (central and/or local jurisdiction) cannot tax-discriminate between the two types of $M_i$s. This is the case because $M_i$'s type is nonverifiable, which ultimately means that a tax-posting regime conditional on types is infeasible because it cannot be enforced in a court of law. In other words, it is not possible for the tax-poster to write in the tax law that the locational tax to be paid by the multinational is conditional on its type, because $M_h$ will pretend to be $M_l$ and there will be no way to prove the contrary in the court. Consequently, the central government can only set taxes conditional on the jurisdiction where $M_i$ builds the new plant, not on $M_i$'s type. Notice that nonverifiability means that $M_i$'s type is not known by the court (or by the tax-poster).

**Figure 1**
*Sequence of Events*

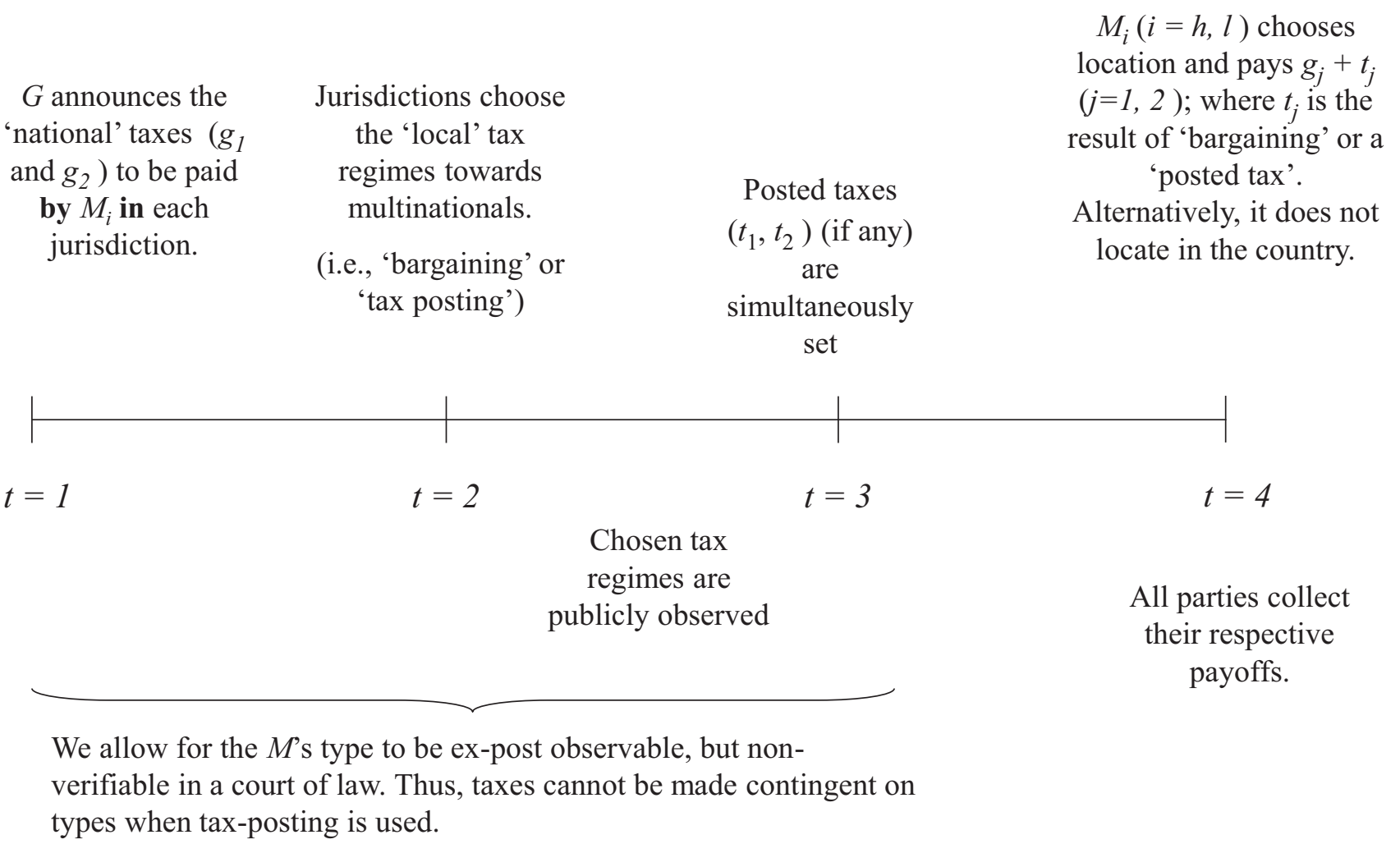

In the second, third, and fourth stages each jurisdiction makes choices in order to maximize its own expected tax collection. In the second stage the two jurisdictions simultaneously choose their local-tax regime – tax-bargaining or tax-posting. In the third stage, when the chosen tax regimes are publicly observed, the jurisdiction that has chosen the tax-posting regime (if any) announces its locational tax; if both jurisdictions have chosen the tax-posting regime, they simultaneously announce their locational taxes, $t_1$ and $t_2$. In the fourth stage $M_i$ shows up and chooses whether to locate the production plant in one of the jurisdictions or not to come to the country at all. In the case that $M_i$ locates in a jurisdiction, it has to pay both the central-government and the jurisdictional taxes. Depending on which tax regime was chosen by the winning jurisdiction in stage 2, this last tax will be a posted tax or the result of a bargaining process. Recall that the tax base is the locational rents produced by MNCs and that profit tax rates are not the appropriate tool to tax them. Thus, each of the jurisdictions will find it optimal to exempt the MNCs from the payment of the profit tax rate and instead demand the payment of a locational tax.

The payoffs for all the players are realized in the fourth stage of the game. When $M_i$ shows up, the jurisdiction with the lowest (highest) aggregate tax $J_j$ ($J_k$) for $j,k = 1,2,\ j \neq k$ – i.e., $g_k + t_{ik} \geq g_j + t_{ij}$ – gets the payoff shown in the following expression:[10,11]

$$w_{ij} = \begin{cases} t_{ij} & \text{if } r_i - g_j - t_{ij} \geq 0, \\ 0 & \text{if } r_i - g_j - t_{ij} < 0, \end{cases} \qquad w_{ik} = 0, \tag{1}$$

where the subscript $i$ represents the fact that, under the tax-bargaining regime, the local tax paid by $M_i$ depends on its type. $M_i$'s payoff and the country's *ex post* welfare are respectively

$$\pi_i = w_i^M = \max\{r_i - g_j - t_{ij}, 0\}, \tag{2}$$

$$W_i = \begin{cases} t_{ij} + g_j & \text{if } r_i - g_j - t_{ij} \geq 0, \\ 0 & \text{if } r_i - g_j - t_{ij} < 0. \end{cases} \tag{3}$$

Given that the probabilities of $M_h$ and $M_l$ showing up are $q$ and $1-q$ respectively, the calculation of the expected jurisdictional payoffs and country's expected welfare is straightforward from (1) and (3).

In order to get the results mentioned in the introduction, it is necessary to find the country's expected welfare and the expected jurisdictional payoffs under different values of the set $(g_1, g_2)$ (first stage of the game). However, we first need to find out whether the jurisdictions choose the tax-bargaining or the tax-posting regime (second stage) as well as their subgame equilibrium taxes and payoffs (third and/or fourth stages). There are three possible subgames:

**Subgame** $(b, p)$ **or** $(p, b)$**:** One jurisdiction commits to tax-posting while the other commits to tax-bargaining.
**Subgame** $(p, p)$**:** Both jurisdictions commit to tax-posting.
**Subgame** $(b, b)$**:** Both jurisdictions commit to tax-bargaining.

We adopt the convention that the first (second) element of a bracket, say $(p, b)$ or $(p, p)$, refers to the favored (nonfavored) jurisdiction. In section 4 the equilibrium jurisdictional payoffs for these three subgames are used to find $G$'s optimal policy in the first stage of the game. The main results of this paper appear under the parameter values satisfying the following assumption:

**Assumption 1** $(1-q)r_l + qr_l + q\frac{r_h - r_l}{2} > qr_h \Leftrightarrow q < \frac{2r_l}{r_h + r_l}$.

10 For simplicity, we are assuming that the jurisdictions do not consider the central-government tax revenue in their own payoff functions. Obviously, this is not necessarily a realistic assumption if the way the central government spends this tax revenue results in higher benefits for the competing jurisdictions. However, one justification for this assumption can be the existence of a large number of jurisdictions in the country. In this case, each jurisdiction would get negligible benefits from this central-government tax revenue. Indeed, the central government could expend this tax revenue in a way that only increases the welfare of the jurisdictions that are not participating in the competition for $M_i$.

11 Notice that we are making use of convenient tie-breaking rules when $r_i - g_j - t_{ij} = 0$. Other tie-breaking rules will be adopted below when needed, but for brevity they will not be made explicit.

The first inequality means that the country's maximum expected welfare from attracting both $M_i$s' types is higher than the maximum from only attracting $M_h$. The left-hand side is the country's expected welfare if (a) the central-government tax on the favored jurisdiction is $g_1 = r_l$, (b) this jurisdiction adopts tax bargaining, and (c) the outside option is not binding.[12] More precisely, the first (second) term inside the first member of the inequality is the expected central-government tax to be levied from the low-(high-)type $M$. The last term inside the first member is the expected favored-jurisdiction tax to be levied from a high-type $M$, which is equal to half of the after-central-government-tax surplus ($\frac{r_h - r_l}{2}$) multiplied by the probability that an $M_h$ shows up. Notice that, because the outside option is binding, the after-central-government-tax surplus is equally divided between the favored jurisdiction and $M_h$.

Assumption 1 simply requires the parameter values to be such that it will never be optimal for the central government to set taxes such that the country attracts only $M_h$. This would be equivalent to disregarding a corner solution in a case with an infinite number of firm types (instead of the two types assumed in this paper). A consequence of assumption 1 is that, in the optimum, at least one of the central-government taxes is not higher than $r_l$; that is, $g_1 \leq r_l$. Additionally, and in order to avoid the analysis of trivial cases, the following simplifying assumption is also made. Not making this assumption would make the proofs unnecessarily longer and less neat.

**Assumption 2** $0 \leq g_1$ and $g_2 \leq r_h$.

On the one hand, the simplifying assumption $0 \leq g_1$ does not change the results, because setting $g_1 < 0$ is clearly a dominated strategy for the country. That is, it would result in a lower welfare for the country than setting $g_1 = 0$ if the favored jurisdiction chooses bargaining. Clearly the central-government revenue will be decreasing in the size of the subsidy, while the favored jurisdiction will recover only a share of this subsidy in the bargaining with $M_i$. Moreover, by setting $g_1 < 0$ the country's welfare will not be higher than by setting $g_1 = 0$ if the favored jurisdiction chooses tax-posting. That is, in the best scenario for the country the favored jurisdiction would increase its tax by the amount of the subsidy. On the other hand, the simplifying assumption $g_2 \leq r_h$ is justified because whether the central-government tax in the nonfavored jurisdiction, $J_2$, is equal to or higher than $r_h$ does not make any

12 Later on we will show that, conditional on both $M_i$s' types being attracted, this scenario is the preferred one for the country. We will also show that, given $g_1 = r_l$ and the central government setting the right tax on the nonfavored jurisdiction (to be determined latter), tax bargaining will be incentive-compatible for the favored jurisdiction. Now, however, we are just assuming that this is the case.

difference in the results. Thus, in both cases $J_2$ will become unattractive for any of the multinationals.

## 3. Jurisdictional Tax Game Equilibria

The following three subsections characterize the subgames' equilibrium expected jurisdictional payoffs, which are then used in section 4 to compute the subgame-perfect equilibrium of the entire game.

### 3.1. Mixed Tax Regime

In this subsection we look at the subgame where (at the second stage) one jurisdiction chooses the tax-bargaining regime and the other the tax-posting one. Thus, in the third stage of the game the tax-poster chooses the particular level of tax to be imposed on $M_i$. Finally, in the fourth stage, $M_i$ shows up and bargains with the tax-bargainer over the tax to be paid in its location.

For the fourth stage of the game we use a standard Rubinstein alternating-offer bargaining game with outside option (Osborne and Rubinstein, 1990).[13] In the particular case where $J_1$ chooses a tax-bargaining regime, the game is one where $J_1$ and $M_i$ bargain over a pie of size $s_{i1} = r_i - g_1$ (called the *surplus*) and where $M_i$ can opt out and get an *outside option* equal to $\max\{s_{i2} - t_2, 0\}$, where $s_{i2} = r_i - g_2$. That is, $M_i$'s outside option is the greater of what $M_i$ obtains by locating in $J_2$ (the tax-poster) and its participation constraint. $J_1$ has no outside option.

Two cases have to be considered:

**(a) Subgame** $(b, p)$**:** a subgame where $J_1$ ($J_2$) chooses the tax-bargaining (tax-posting) regime. The equilibrium jurisdictional taxes are

$$t_{i1}^* = \min\left\{\frac{s_{i1}}{2}, t_2^* + g_2 - g_1\right\}, \quad t_2^* \geq 0. \tag{4}$$

Under Rubinstein bargaining, the two bargainers would strike a deal if $M_i$'s surplus in $J_1$, $r_i - g_1$, is not negative and not lower than $M_i$'s surplus in $J_2$, $r_i - g_2 - t_2$. Given $g_1 \leq g_2$, $g_1 \leq r_l$, and $t_2 \geq 0$, these two conditions are satisfied in our case; hence $J_1$ will always undercut $J_2$ and attract both $M_i$'s types. Then, the tax $t_{i1}^*$ in (4) results from the fact that under Rubinstein

13 As is clearly shown by Binmore et al. (1989), the axiomatic Nash bargaining solution does not provide the right way to deal with outside options. They rightly claim that the way to deal with outside options in bargaining is by applying an analysis of optimal strategic behavior in a game-theoretic model like the one proposed by Osborne and Rubinstein (1990). Besides, this bargaining model provides the same prediction as the Bolton and Whinston (1993) three-party bargaining game used in section 3.3.

**Table 2**
*Equilibrium Expected Jurisdictional Payoffs for each of the Subgames.*

| | |
|---|---|
| **1** | $\begin{pmatrix} w_1^*(b,p) \\ w_2^*(b,p) \end{pmatrix} = \begin{pmatrix} q\min\left\{\frac{s_{h1}}{2}, g_2 - g_1 + t_2^*\right\} + (1-q)\min\left\{\frac{s_{l1}}{2}, g_2 - g_1 + t_2^*\right\} \\ 0 \end{pmatrix}$ |
| **2** | $\begin{pmatrix} w_1^*(p,b) \\ w_2^*(p,b) \end{pmatrix} = \begin{cases} \begin{pmatrix} g_2 - g_1 \\ 0 \end{pmatrix} & \text{if } g_1 < g_2 \leq r_l < r_h \quad \textbf{(a)} \\ \begin{pmatrix} \max\{s_{l1}; q(g_2 - g_1)\} \\ 0 \end{pmatrix} & \text{if } g_1 \leq r_l < g_2 \leq r_h \quad \textbf{(b)} \end{cases}$ |
| **3** | $\begin{pmatrix} w_1^*(p,p) \\ w_2^*(p,p) \end{pmatrix} = \begin{cases} \begin{pmatrix} g_2 - g_1 \\ 0 \end{pmatrix} & \text{if } g_1 \leq g_2 \leq r_l < r_h \quad \textbf{(a)} \\ \begin{pmatrix} \max\{s_{l1}; q(g_2 - g_1)\} \\ 0 \end{pmatrix} & \text{if } g_1 \leq r_l < g_2 \leq r_h \quad \textbf{(b)} \end{cases}$ |
| **4** | $\begin{pmatrix} w_1^*(b,b) \\ w_2^*(b,b) \end{pmatrix} = \begin{pmatrix} q\min\left\{\frac{s_{h1}}{2}, g_2 - g_1\right\} + (1-q)\min\left\{\frac{s_{l1}}{2}, g_2 - g_1\right\} \\ 0 \end{pmatrix}$ |

Note: $g_1 \leq r_l$ (from assumption 1), $0 \leq g_1$ and $g_2 \leq r_h$ (from assumption 2), and $t_2^* \geq 0$ (from (4)).

bargaining $M_i$ (the bargainer with the outside option) gets a payoff equal to $\max\left\{\frac{s_{i1}}{2}, s_{i2} - t_2^*\right\}$; i.e., the greater of half the surplus and the value of its outside option. The second term inside the minimum operator in (4) can be interpreted as being equal to $J_2$'s quoted tax, $t_2^*$, adjusted by $J_2$'s competitive disadvantage, $g_2 - g_1$. On the other hand, the tax $t_2^*$ in (4) results from the fact that $J_2$ cannot attract $M_i$ even on setting $t_2 = 0$. Moreover, given that $J_1$ plays after $J_2$, the latter would also get a payoff of zero on setting $t_2 > 0$.

The multiple equilibria, $t_2^* \geq 0$, are payoff-equivalent for $J_2$, but not for $J_1$. These multiple equilibria are similar to the ones that would appear in a homogeneous-product and asymmetric-costs Stackelberg–Bertrand duopoly. However, even though there are multiple equilibria in this subgame, there will be a unique equilibrium for the whole game, as will be shown in section 4. Finally, using (4) and (1), the corresponding equilibrium expected jurisdictional payoffs can be obtained; they are reported in row 1 of table 2.

**(b) Subgame** $(p,b)$**:** a subgame where $J_1$ ($J_2$) chooses the tax-posting (tax-bargaining) regime. The equilibrium jurisdictional taxes are[14]

$$\begin{pmatrix} t_1^* \\ t_{i2}^* \end{pmatrix} = \begin{pmatrix} g_2 - g_1 \\ 0 \end{pmatrix} \text{ if } g_1 < g_2 \leq r_l < r_h \text{ and} \tag{5a}$$

$$\begin{pmatrix} t_1^* \\ t_{i2}^* \end{pmatrix} = \begin{pmatrix} \begin{cases} s_{l1} & \text{if } s_{l1} \geq q\,(g_2 - g_1) \\ g_2 - g_1 & \text{if } s_{l1} < q\,(g_2 - g_1) \end{cases} \\ 0 \end{pmatrix} \text{ if } g_1 \leq r_l < g_2 \leq r_h, \tag{5b}$$

where $s_{l1} = r_l - g_1$. On the one hand, in its frustrated attempt to attract an $M_i$, $J_2$ will reduce its tax to zero. On the other hand, when trying to attract a particular $M_i$, $J_1$ faces two restrictions on the level of its tax: the competition from $J_2$ ($g_1 + t_1^* \leq g_2 + t_{i2}^*$) and $M_i$'s participation constraint ($g_1 + t_1^* \leq r_i$). When the former is more restrictive than the latter, $M_i$ would locate in $J_1$ if its tax is not higher than $J_2$'s competitive disadvantage, $g_2 - g_1$ (a tie-breaking rule is in place).

Given $g_1 < g_2$, $g_1 \leq r_l$, and $g_2 \leq r_h$, the only two possible scenarios are the ones determined by the conditions in (5a) ($g_1 < g_2 \leq r_l < r_h$) and (5b) ($g_1 \leq r_l < g_2 \leq r_h$). When the conditions in (5a) apply, the competition from $J_2$ is more restrictive than either $M_i$'s participation constraint. Then, $J_1$'s optimum tax is $t_1^* = g_2 - g_1$, which attracts both $M_i$.

On the contrary, in (5b) the main restriction for $J_1$, when willing to attract $M_l$ ($M_h$), is $M_l$'s participation constraint (the competition from $J_2$); hence it will maximize its expected payoff subject to the following two options: (a) the setting of a tax that just satisfies $M_l$'s participation constraint, $t_1^* = s_{l1}$, which both attracts $M_i$ and provides $J_1$ an expected payoff of $w_1^*(p,b) = s_{l1}$, or (b) the setting of a tax that just beats $J_2$, $t_1^* = g_2 - g_1$, but that is only accepted by $M_h$ and results in $w_1^*(p,b) = q(g_2 - g_1)$. Finally, using (5) and (1), the corresponding equilibrium expected jurisdictional payoffs can be obtained; they are reported in row 2 of table 2.

### 3.2. Tax-Posting Regime

We now look at the case where in the second stage of the game both jurisdictions have already committed to implement the tax-posting regime – subgame $(p,p)$. In the third stage of the game the local jurisdictions simultaneously announce nonnegotiable taxes. This continuation game entails Bertrand-type

14 Notice that we now have only $g_1 < g_2$, because $g_1 = g_2$ was already considered in the subgame $(b,p)$.

tax competition. The equilibrium jurisdictional taxes are

$$\begin{pmatrix} t_1^* \\ t_2^* \end{pmatrix} = \begin{pmatrix} g_2 - g_1 \\ 0 \end{pmatrix} \text{ if } g_1 \leq g_2 \leq r_l < r_h \quad \text{and} \tag{6a}$$

$$\begin{pmatrix} t_1^* \\ t_2^* \end{pmatrix} = \begin{pmatrix} \begin{cases} s_{l1} & \text{if } s_{l1} \geq q(g_2 - g_1) \\ g_2 - g_1 & \text{if } s_{l1} < q(g_2 - g_1) \end{cases} \\ 0 \end{pmatrix} \text{ if } g_1 \leq r_l < g_2 \leq r_h, \tag{6b}$$

and the equilibrium expected jurisdictional payoffs are the ones reported in row 3 of table 2. Not surprisingly, the results of subgames $(p,p)$ and $(p,b)$ are identical. This is the case because whether the nonfavored jurisdiction implements a tax-posting or a tax-bargaining regime does not change the fact that in the frustrated attempt to attract a multinational it reduces its tax to zero. Given that, the result should also be the same for the favored jurisdiction, because in both cases it is a tax-poster.

Proof. See appendix 6.1.

### 3.3. Tax-Bargaining Regime

We now look at the case where in the second stage of the game both jurisdictions have already committed to implement the tax-bargaining regime – subgame $(b,b)$. To model this negotiation process we adopt the noncooperative three-party bargaining game developed by Bolton and Whinston (1993). In our context, this is an alternating-offer game where $M_i$ has to make offers to the two jurisdictions. The unique equilibrium outcome of the Bolton and Whinston model is:

**Lemma 1** Agreement is immediate, $M_i$ never takes its outside option, and its payoff is the greater of:

1. half of the surplus it creates in the favored jurisdiction, $\frac{s_{i1}}{2}$, and
2. $M_i$'s outside option, which is the surplus it creates in the nonfavored jurisdiction, $s_{i2}$.

Proof. See Bolton and Whinston (1993).

The following lemma expresses the results of this bargaining game in terms of the expected jurisdictional payoffs.

**Lemma 2** Given $g_1 \leq g_2$ as well as $g_1 \leq r_l$ and $g_2 \leq r_h$ (from assumptions 1 and 2), the equilibrium expected jurisdictional payoffs of the subgame are the ones reported in row 4 of table 2.

**Proof.** The proof is straightforward from lemma 1. When $M_i$ gets $s_{i1}/2$, $J_1$ also gets $s_{i1}/2$, and when $M_i$ gets $s_{i2}$, $J_1$ is the residual claimant and gets $s_{i1} - s_{i2} = g_2 - g_1$. ■

## 4. Country's Optimal Policy

This section analyzes the optimal policy for the country. The determination of the central government's optimal taxes is done in the next subsection. Section 4.2 performs a country's welfare comparison under different scenarios of tax-related jurisdictional autonomy as well as under complete centralization. Finally, section 4.3 discusses the similarities and differences with the literature.

### 4.1. Central-Government Optimal Taxation

In the first stage of the game the central government has to find the set $(g_1, g_2)$ that maximizes the country's welfare. Therefore, we first need to find the expected country's welfare associated with each of table 2's jurisdictional payoff vectors. This is done in table 3 by using (3) and the fact that $M_h$ and $M_l$ show up with probabilities $q$ and $1-q$ respectively.

In the first stage of the game the central government should maximize the expected country's welfare restricted to each of the four subgames being the equilibrium of the whole game. The highest country's welfare among these

**Table 3**
*Equilibrium Expected Country's Welfare for each of the Subgames.*

| | |
|---|---|
| **1** | $W^*(b,p) = q\min\left\{\frac{r_h+g_1}{2}, g_2+t_2^*\right\} + (1-q)\min\left\{\frac{r_l+g_1}{2}, g_2+t_2^*\right\}$ |
| **2** | $W^*(p,b) = \begin{cases} g_2 & \text{if } g_1 < g_2 \le r_l < r_h \quad \textbf{(a)} \\ \max\{r_l; qg_2\} & \text{if } g_1 \le r_l < g_2 \le r_h \quad \textbf{(b)} \end{cases}$ |
| **3** | $W^*(p,p) = \begin{cases} g_2 & \text{if } g_1 \le g_2 \le r_l < r_h \quad \textbf{(a)} \\ \max\{r_l; qg_2\} & \text{if } g_1 \le r_l < g_2 \le r_h \quad \textbf{(b)} \end{cases}$ |
| **4** | $W^*(b,b) = q\min\left\{\frac{r_h+g_1}{2}, g_2\right\} + (1-q)\min\left\{\frac{r_l+g_1}{2}, g_2\right\}$ |

Note: $g_1 \le r_l$ (from assumption 1), $0 \le g_1$ and $g_2 \le r_h$ (from assumption 2), and $t_2^* \ge 0$ (from (4)).

four restricted ones would be the unrestricted maximum welfare, and the set of central-government taxes producing it would be the optimal one.

However, table 2 shows that $J_2$ is indifferent between the tax-posting and the tax-bargaining regimes (it always gets a payoff of zero); hence it would choose each of these regimes with some positive probability. Yet, the country's welfare is independent of the regime adopted by $J_2$.[15] Thus, we can ignore $J_2$'s choice and look at two (instead of four) restricted maximization problems – that is, when the country's welfare is maximized subject to the set $(g_1, g_2)$ being incentive-compatible with $J_1$ respectively choosing bargaining and tax-posting.

Thus, on the one hand the set of optimal taxes conditional on $J_1$ choosing bargaining is

$$(g_1, g_2) = \left(r_l, \frac{r_h + g_1}{2}\right), \tag{7}$$

which results in a country's welfare

$$W^e(b, \cdot) = q\frac{r_h + r_l}{2} + (1-q)r_l. \tag{8}$$

This is the case because the country's welfare, incentive-compatible with $J_1$ choosing bargaining (from table 2), is weakly increasing in both $g_1$ and $g_2$ until $(g_1, g_2) = \left(r_l, \frac{r_h+g_1}{2}\right)$ (from rows 1 and 4 of table 3). Notice that the set $(g_1 = r_l,\ g_2 > \frac{r_h+g_1}{2})$ is not compatible with $J_1$ choosing bargaining, because $J_1$ would get a higher expected payoff by posting a tax $t_1 = g_2 - g_1$ that only attracts $M_h$ (from table 2) – $J_1$'s expected payoff from implementing bargaining would be $q\frac{r_h - r_l}{2}$, which is lower than $q(g_2 - g_1)$. A similar reasoning applies if $g_1 > r_l$.

On the other hand, it is clear from rows 2 and 3 of table 3 that the country's welfare incentive-compatible with $J_1$ choosing tax-posting cannot be higher than $\max\{r_l, qr_h\}$, which, given assumption 1, is lower than the value in (8). Thus, the (unrestricted) optimal central-government policy is the one in (7).

A clear first implication of the result in (7) is, as in Parcero (2007), the optimality of the asymmetric central-government tax treatment of the two identical jurisdictions. By setting a higher tax in one of the jurisdictions, the central government increases the bargaining power of the other jurisdiction vis-à-vis the multinational. A second, less obvious implication is the following:

**Proposition 1** Some degree of competition between the jurisdictions is desirable. However, this is not required if the local jurisdictions can only apply tax bargaining – i.e., the tax-posting strategy is banned.

15 Strictly speaking, it does depend on $J_2$'s choice in the case that $J_1$ implements the bargaining regime (contrast rows 1 and 4 of table 3), but not in the vicinity of the optimum; this will become clearer in the analysis that follows.

This result is a consequence of the existence of an out-of-equilibrium conflict of interest between the favored jurisdiction and the central government, which can also be understood as the favored jurisdiction being willing to set a tax higher than what would be optimal for the country. In particular, given $g_1 = r_l$, a too high tax in the nonfavored jurisdiction (i.e., $g_2 > \frac{r_h+r_l}{2}$) will lead the favored jurisdiction to prefer the implementation of a tax-posting strategy (with a high tax targeting $M_h$), whereas the whole country would prefer the latter to implement tax-bargaining. Notice that the favored jurisdiction would not get any benefit from attracting $M_l$; the tax that this multinational pays to the central government does not leave margin for any additional tax. Thus, the favored jurisdiction would do better by implementing a posted tax that extracts $M_h$'s entire after-central-government-tax surplus, $r_h - g_1$, which is higher than what this jurisdiction could obtain by implementing a bargaining tax, $\frac{r_h-g_1}{2}$. Clearly, an alternative way of solving the conflict of interest would be to ban the local jurisdiction's adoption of the the tax-posting strategy.

### 4.2. Country's Welfare under Different Scenarios

Let us now investigate which one of the following three scenarios is preferred in terms of the country's welfare. *Scenario* $(i)$: The jurisdictions of the country can choose between the implementation of the tax-bargaining and the tax-posting regimes. *Scenario* $(ii)$: The jurisdictions of the country can only implement the tax-bargaining regime. *Scenario* $(iii)$: The jurisdictions of the country can only implement the tax-posting regime. We get the following results:

**Proposition 2** First, scenarios (i) and (ii) provide the same equilibrium expected welfare of the country; i.e., $W^*_{(i\&ii)} = \max\left\{q\frac{r_h+r_l}{2} + (1-q)r_l, qr_h\right\}$. Second, when the attraction of only $M_h$ is the country's optimal policy in each of the three scenarios (in this case assumption 1 does not apply), they produce the same expected welfare of the country; i.e., $W = qr_h$. Third, when the attraction of both $M_i$'s types is the optimal country's policy under scenarios (i) and (ii) (in which case it may or may not be also the optimal one under scenario (iii)), scenarios (i) and (ii) would strictly welfare-dominate scenario (iii); i.e., $W^*_{(i\&ii)} = q\frac{r_h+r_l}{2} + (1-q)r_l > W^*_{(iii)} = \max\{r_l, qr_h\}$. Thus, we can conclude that scenarios (i) and (ii) weakly welfare-dominate scenario (iii).

Proof. See appendix 6.2.

From the previous proposition we get that scenarios (i) and (ii) provide the same country's welfare. However, it is worth mentioning that the implementation of each of them has its own difficulties. On the one hand, scenario (ii) requires the banning of the local jurisdictions' tax-posting strategy, which may not be easily accomplished. On the other hand, in scenario (i) the conflict of interest between the central government and the favored jurisdiction requires a careful calibration of the tax in the nonfavored jurisdiction, $g_2 = \frac{r_h + g_1}{2}$, in order for the competition exerted by the latter to be neither too high nor too low. If the central government gets this fine-tuning wrong, the country's welfare may be lower than the value in (10).

Assume now that a country is characterized by scenario (i) and that it can decide whether or not to restrict the local jurisdictions' ability of policymaking. This may be the case with respect to the time of a constitutional change, the passage of a federal law, or the establishment of a new SEZ system. Then, the following corollary applies:

**Corollary 1** It is not optimal to restrict the ability of jurisdictions to choose the tax-bargaining regime.

This policy recommendation can be explained as follows. When the favored jurisdiction tax-posts, the country as a whole is implementing a tax-posting regime (i.e., the addition of the two levels of government's posted taxes is equivalent to one posted tax). On the contrary, when the favored jurisdiction implements bargaining, the country as a whole is implementing a bargaining-with-reservation-tax regime (the reservation tax being the central-government tax in the favored jurisdiction), which, under the optimal central-government tax, provides a higher welfare of the country (aggregate revenue) than a tax-posting regime (see footnote 3).

Let us now analyze the country's welfare under two alternative scenarios. First, a scenario where the taxation of multinationals rents is the exclusive responsibility of the local jurisdictions. This case clearly results in a *race to the bottom* – the two identical local jurisdictions set a tax of zero and generate zero welfare for the country. Second, a scenario where the taxation of MNCs is the exclusive responsibility of the central government (a move from a federal to a unitary system if the jurisdictions are states). Given that the central government posts its taxes, this last scenario provides the same country welfare as scenario (iii), which is welfare-dominated, as per the argument in the previous paragraph. Moreover, this second scenario will also be welfare-dominated if the central government implements tax-bargaining instead of tax-posting. In this case, the country's welfare will be lower than under scenario (ii). Under scenario (ii) the country is implementing a bargaining-with-reservation-tax regime, which, as we explained in footnote 3, provides a higher country welfare than a simple bargaining regime. Thus, the following corollary applies:

**Corollary 2** A scenario in which both central and local governments set taxes welfare-dominates a scenario where there are only central-government taxes or only local-government taxes.

### 4.3. Relation to Similar Results in the Literature

Our results are related to findings in the concurrent-taxation literature – in particular, the mentioned out-of-equilibrium conflict of interest and corollary 1's result that the taxation of MNCs should be the joint responsibility of both the central government and local jurisdictions. The similarities and differences are discussed below.

First, and as already mentioned, the out-of-equilibrium conflict of interest can also be read as the favored jurisdiction being willing to set a tax higher than what would be optimal for the country. This is the case because the two levels of governments share the same tax base, but the local jurisdiction does not take the tax collected by central government into account when choosing its tax strategy. That is, the local jurisdiction produces a negative fiscal externality to the central government, leading to a jurisdictional tax higher than what would be optimal for the country.

A similar vertical externality is found in the concurrent-taxation literature (e.g., Keen and Kotsogiannis, 2002), though only local firms are considered and the tax is on capital. In this literature, the welfare loss is produced by a misallocation of resources – e.g., an overprovision (underprovision) of the local (the central) government's public good in Keen and Kotsogiannis (2002). On the other hand, in our paper the country's welfare loss comes from the low appropriation of foreign firms' locational rents – a purely tax-revenue matter.

The vertical-externality problem is solved by allowing some degree of tax competition between local jurisdictions. This solution is common to both our paper and the concurrent-taxation literature. Besides, in our model the problem is also avoided if the local jurisdictions are only able to apply the bargaining strategy. The latter case is the one considered in Parcero (2007), in which the vertical-externality problem is not present. This shows that in allowing the jurisdictions to choose between bargaining and tax-posting, this paper is an appropriate extension to Parcero (2007).

Second, corollary 1's result is a contribution to the debate on centralized versus decentralized taxation, which has been also investigated in the context of the concurrent-taxation literature. In this approach, the optimal degree of decentralization is the result of a trade-off between inefficiencies at the local and central government levels. The former are inefficiently low capital taxes and public-good provision, which are typically a result of a race to the bottom in the local governments' competition for scarce capital (Wilson and Janeba, 2005; Janeba and Wilson, 2011) or the existence of other interregional externalities (Janeba and Wilson, 1999). On the other hand, the central-government inefficiencies have been modeled as produced by a uniform provision of public goods (Koethenburger, 2008) or by an unequal and inefficient division of public-good expenditures across localities (Janeba and Wilson, 2011).

However, our results do not suggest an optimal degree of decentralization, but that the taxation should be the joint responsibility of both central government and local jurisdictions. This taxation, which in our framework is on foreign firms' locational rents rather than on perfectly competitive local firms' capital, takes the form of an aggregate "pricing" strategy to appropriate these locational rents. Complete centralization is not optimal, because it results in a lower bargaining power of the country than under coexistence of two levels of jurisdiction. On the other hand, complete decentralization results in no extraction of the MNCs' locational rents, because of the race to the bottom in the tax competition between the local jurisdictions. Thus, the optimum is achieved when there is coexistence of two levels of jurisdictions; but this involves the solution of two problems: first, the race to the bottom between local jurisdictions, and second, the fact that a shared tax base between the central government and the favored local jurisdiction leads to negative vertical fiscal externalities if this jurisdiction is able to adopt a tax-posting strategy and there is no competition from other jurisdictions.

## 5. Conclusion

This paper looks at a country's optimal central-government policy in a setting where its local jurisdictions compete for the attraction of footloose multinationals to their sites. It is original in the modeling of some stylized facts that have been somewhat overlooked by the literature. First, it allows locational rents to be different from profits, which is a very realistic feature that only by chance would not apply. Second, the local jurisdictions are allowed to choose between the implementation of firm-specific and non-firm-specific strategies. Third, it captures a distinctive characteristic of firm-specific strategies – viz., the existence of a negotiation with the MNC. Naturally, this negotiation is done at the time the MNC shows up, in contrast to the in-advance setting of taxes in the tax-posting regime. The previous literature neither discussed an alternative non-firm-specific strategy (e.g., Parcero, 2007), nor made a distinction according to timing – i.e., both types of policies were allowed to be set either in advance or at the time the MNC shows up (see Amerighi and De Feo, 2014, for an example of the former).

The following results are found. First, as in Parcero (2007), we get that the central government's asymmetric tax treatment of the two identical jurisdictions welfare-dominates the symmetric one. This asymmetric treatment increases the bargaining power of the low-tax jurisdiction vis-à-vis the multinational one. The central government's decision to privilege one jurisdiction relative to others is based on efficiency grounds, but we acknowledge the fact that in real life this decision is sometimes taken on the grounds of achieving a more equal development of the country's jurisdictions. However, our paper's advice of asymmetric jurisdictional treatment by the central government does not necessarily result in more unequal jurisdictional development. First, it may be that on efficiency grounds it is better to give privileges to a less-developed jurisdiction. Second, it may be that the central government gives special privileges to one jurisdiction in order to attract firms belonging to a particular industry while at the same time giving special privileges to another jurisdiction in order to attract firms of another industry. Evidence in favor of this second point might be the central government's creation of SEZs that are located in different jurisdictions and with the aim of targeting different industries.

Second, we find that some jurisdictional tax competition is desirable if the local jurisdictions have the ability to choose between the tax-bargaining and a tax-posting strategy. In other words, the central government's locational tax in the nonfavored jurisdiction should not be too high. This is the case because a certain degree of competition from the nonfavored jurisdiction drives the favored one to adopt the tax-bargaining regime, which is the optimal one for the country.

Third, and clearly related to the previous result, we show that the implementation of the jurisdictional tax-bargaining regime weakly welfare-dominates the implementation of the jurisdictional tax-posting regime. Consequently, if indeed our choice of proxies for the policy's firm-specificity measures is an appropriate simplification of the reality, our paper's advice would be against the banning of the jurisdictional implementation of firm-specific policies. Moreover, if a country decides to (and has the capacity to) restrict the jurisdictional ability of policymaking in some way, our paper would advocate the restriction of the non-firm-specific policies, rather than the firm-specific ones. Fourth, our results suggest that a higher extraction of the multinational's locational rents will be achieved if their taxation is the joint responsibility of both the central government and the local jurisdictions.

Finally, in this paper the firm-specific and non-firm-specific strategies were respectively characterized by variants of bargaining and tax-posting. However, other alternatives should also be explored in future research. For instance, the consideration of some of the typical assumptions and frameworks of the literature on price-bargaining versus price-posting literature, which in our context could be translated as (a) jurisdictional heterogeneity; (b) interjurisdictional externalities; (c) private information in terms of jurisdictional characteristics and/or valuations of the multinationals; and (d) different versions of search models (directed versus random search). Another alternative could be to analyze the policy's firm-specificity issue with a mix of jurisdictional infrastructure provision and taxation – that is, of whether the infrastructure is built in advance or tailor-made after a particular multinational shows up. This alterna-tive approach would have the advantage of increasing and taxing (instead of just taxing) the rents created by the multinational(s).

## 6. Appendix

### 6.1. Proof for the Tax-Posting Regime's Equilibrium

Notice that whichever jurisdiction has a higher aggregate tax would not attract any $M_i$; hence in its frustrated attempt to attract an $M_i$, $J_2$ will reduce its tax to zero. The rest of the proof is exactly the same as the proof for the subgame $(p,b)$, with the difference that we now have $g_1 \leq g_2$ instead of $g_1 < g_2$, $t_2^*$ instead of $t_{i2}^*$, we refer to 6a (6b) instead of 5a (5b), and the jurisdictional payoff is $w_1^*(p,p)$ instead of $w_1^*(p,b)$. ■

### 6.2. Proof of Proposition 2

First, notice that whether scenario (i), (ii), or (iii) applies, if assumption 1 does not apply, the country's maximum expected welfare restricted to only $M_h$ being attracted is

$$W_{(h)} = qr_h, \tag{9}$$

which can be achieved by setting $g_2 = g_1 = r_h$.

**Country's equilibrium expected welfare under scenario (i):** We know from section 4.1 that the country's maximum expected welfare under scenario (i), when it is optimal for the country to attract both $M_i$'s types (i.e., assumption 1 applies), is the one in (7). This, together with (9), results in the country's equilibrium expected welfare under scenario (i) as

$$W^*_{(i\&ii)} = \max\left\{q\frac{r_h + r_l}{2} + (1-q)r_l, qr_h\right\}. \tag{10}$$

**Country's equilibrium expected welfare under scenario (ii):** The country's expected welfare under scenario (ii), when it is optimal for the country to attract both $M_i$'s types (i.e., assumption 1 applies), is the one in row 4 of table 3. In this case the central government's optimal taxes are $g_1 = r_l$ and $g_2 \geq \frac{r_h+g_1}{2}$. (This is because the country's expected welfare in row 4 of table 3 is nondecreasing in $g_2$, and, for any $g_1 < r_l$ and $g_2 \geq \frac{r_h+g_1}{2}$, it is strictly increasing in $g_1$.) Thus, the country's maximum expected welfare is $W = q\frac{r_h+r_l}{2} + (1-q)r_l$, which is equal to the value in (7). This, together with (9), results in the country's equilibrium expected welfare under scenario (ii) being equal to the one in (10). Thus, the first statement of the proposition is proved.

**Country's equilibrium expected welfare under scenario (iii):** The country's expected welfare attainable under scenario (iii) is the one in row 3 of table 3. Thus, given assumption 2, the optimal central-government taxes are $0 \leq g_1 \leq r_l$ and $r_l < g_2 \leq r_h$, which results in

$$W^*_{(iii)} = \max\{r_l, qr_h\}. \tag{11}$$

Finally, to prove the second and third statements of the proposition we compare the country's equilibrium expected welfare in scenarios (i) and (ii) with the one in scenario (iii). On the one hand, it is clear from the second term inside the maximum operators in (10) and (11) that, when the attraction of only $M_h$ is the country's optimal policy in each of the three scenarios, they produce the same expected welfare for the country. On the other hand, when the attraction of both $M_i$'s types is the country's optimal policy under scenarios (i) and (ii) (i.e., the first term inside the maximum operator in (10) is higher than the second one), both scenarios would *strictly* welfare-dominate scenario (iii). This would be the case whether, under scenario (iii), it is optimal to attract both $M_i$'s types or only $M_h$. Thus, scenario (iii) is weakly welfare-dominated by scenarios (i) and (ii). ■